\documentclass[%
superscriptaddress,
twocolumn,
reprint,
a4paper,
amsmath,amssymb,
aps,
prd,
]{revtex4-1}

\usepackage{dcolumn}
\usepackage{bm}
\usepackage{graphicx}
\usepackage[colorlinks, linkcolor=blue, anchorcolor=blue, citecolor=blue]{hyperref}
\usepackage{subfigure}
\usepackage{multirow}
\usepackage{acronym}
\usepackage{tabularx}

\usepackage{color}

\begin{document}

\title{Rapid search for massive black hole binary coalescences using deep learning}

\author{Wen-Hong Ruan}
\email{ruanwenhong@ucas.ac.cn}
\thanks{Equal Contribution}
\affiliation{School of Fundamental Physics and Mathematical Sciences, Hangzhou Institute for Advanced Study, University of Chinese Academy of Sciences, Hangzhou 310024, China}
\affiliation{School of Physical Sciences, University of Chinese Academy of Sciences, No.19A Yuquan Road, Beijing 100049, China}
\affiliation{CAS Key Laboratory of Theoretical Physics, Institute of Theoretical Physics, Chinese Academy of Sciences, Beijing 100190, China}
\author{He Wang}
\email{hewang@ucas.ac.cn}
\thanks{Equal Contribution}
\affiliation{International Centre for Theoretical Physics Asia-Pacific, University of Chinese Academy of Sciences, 100190 Beijing, China}
\affiliation{Taiji Laboratory for Gravitational Wave Universe, University of Chinese Academy of Sciences, 100049 Beijing, China}
\affiliation{CAS Key Laboratory of Theoretical Physics, Institute of Theoretical Physics, Chinese Academy of Sciences, Beijing 100190, China}
\author{Chang Liu}
\email{liuchang@ucas.ac.cn}
\affiliation{School of Fundamental Physics and Mathematical Sciences, Hangzhou Institute for Advanced Study, University of Chinese Academy of Sciences, Hangzhou 310024, China}
\affiliation{School of Physical Sciences, University of Chinese Academy of Sciences, No.19A Yuquan Road, Beijing 100049, China}
\affiliation{CAS Key Laboratory of Theoretical Physics, Institute of Theoretical Physics, Chinese Academy of Sciences, Beijing 100190, China}
\author{Zong-Kuan Guo}
\email{guozk@itp.ac.cn}
\affiliation{CAS Key Laboratory of Theoretical Physics, Institute of Theoretical Physics, Chinese Academy of Sciences, Beijing 100190, China}
\affiliation{School of Physical Sciences, University of Chinese Academy of Sciences, No.19A Yuquan Road, Beijing 100049, China}
\affiliation{School of Fundamental Physics and Mathematical Sciences, Hangzhou Institute for Advanced Study, University of Chinese Academy of Sciences, Hangzhou 310024, China}

\begin{abstract}
The coalescences of massive black hole binaries are one of the main targets of space-based gravitational wave observatories.
Such gravitational wave sources are expected to be accompanied by electromagnetic emissions.
Low latency detection of the massive black hole mergers provides a start point for a global-fit analysis to explore the large parameter space of signals simultaneously being present in the data but at great computational cost.
To alleviate this issue, we present a deep learning method for rapidly searching for signals of massive black hole binaries in gravitational wave data.
Our model is capable of processing a year of data, simulated from the LISA data challenge, in only several seconds, while identifying all coalescences of massive black hole binaries with no false alarms.
We further demonstrate that the model shows robust resistance to a wide range of generalization cases, including various waveform families and updated instrumental configurations.
This method offers an effective approach that combines advances in artificial intelligence to open a new pathway for space-based gravitational wave observations.
\end{abstract}

\maketitle

\section{Introduction}

Assessing the ubiquity of massive black holes in earlier times~\cite{Kormendy:1995er, Magorrian_1998}, a large number of massive black hole binaries (MBHBs) are expected to have formed over the course of cosmic history~\cite{1980BegelmanMassiveblackhole}.
The coalescences of MBHBs with total masses of $10^{5}M_{\odot} - 10^{8}M_{\odot}$ due to galaxy mergers provide a primary source of low-frequency gravitational waves (GWs) detectable by the proposed the Laser Interferometry Space Antenna (LISA) observatory~\cite{amaro2017laser}.
Such systems may also produce detectable electromagnetic (EM) emission, which allows us to witness the formation of a quasar following the final merger~\cite{amaro2017laser}.
Furthermore, a recent study found that the EM emission from electrons accelerated at the external forward shock may emerge days to months after the coalescence~\cite{yuan2021post}.
However, the GW spectrum at millihertz frequencies is expected to be populated by many additional sources, including tens of millions of Galactic binaries, thousands of extreme mass ratio binaries and stellar mass black hole binaries~\cite{amaro2017laser} within the sensitivity band of a LISA-like detector.
Therefore, the resulting superposition of GW signals from numerous resolved and unresolved sources needs to be dealt with and poses a great challenge to individually extracting information on every source.

Typically, a global-fit analysis~\cite{PhysRevD.72.043005, Littenberg:2020bxy, littenberg2023prototype} has been considered for exploring the large parameter space of all overlapping and resolvable signals but at great computational cost.
Although GW signals produced by inspiraling MBHBs can stay in-band for months or even years, MBHBs are relatively easy to detect within a short duration due to their accumulating signal-to-noise ratios.
Accordingly, once an MBHB merger is spotted, global-fit analysis can tackle the case with a fixed number of parameters.
In this case, the follow-up parameter estimation can be accelerated around the alert, which is important for determining the sky location of MBHBs in a short time.

In this work, we make the initial attempt to search for MBHB coalescences in LISA data using the deep learning approach~\cite{lecun2015deep}, which has recently gained popularity in the GW community but with the majority of efforts on the ground-based GW data analysis~\cite{cuoco2020enhancing}.
Different from the case in LIGO-Virgo, the characteristics of LISA data are significantly more complicated due to modulation by the motion of detectors~\cite{Rubbo:2003ap}.
The time-delay interferometry (TDI) technique ~\cite{Tinto:2004wu} involved further increases the complexity of the data for suppressing the laser frequency noise.
Currently, only a few works using neural networks have been done on fast waveform modelling~\cite{Chua:2018woh,Chua:2020stf} and Bayesian inference~\cite{Chua:2019wwt} for GW sources in LISA band.
In this case, it is worth testing the capability of deep learning on realistic LISA data for GW detection.

For our attempt, we construct a matched-filtering convolutional neural network (MFCNN)~\cite{PhysRevD.101.104003} to distinguish whether there are MBHB mergers in a data segment.
Here, we use the model to identify the short data segments containing MBHB mergers from long-duration data.
The coalescence time is limited to a small range and the number of mergers can be determined.
Valuable as a form of data pre-processing, this greatly reduces the computational cost of follow-up parameter estimation (including sky location).
Our model takes only several seconds to analyse a year of LISA data with no false alarms.
Moreover, all MBHB coalescences are identified within several days of data.
It also shows robustness against various waveform families and generations of TDI techniques.
Thus, our model is expected to deal with realistic LISA data and be easily extended to other space-based GW detectors such as Taiji~\cite{Ruan:2018tsw}.

This paper first proceeds with a thorough description of the proposed approach as presented in Sec.~\ref{sec:model}.
Next, Sec.~\ref{sec:strategy} reviews the commonly used strategies in searching and introduces the search methodology on streaming data used in this work.
In Sec.~\ref{sec:Training}, we describe the training datasets and implementation details.
The results are presented in Sec.~\ref{sec:result}.
Finally, summary and discussions are provided in Sec.~\ref{sec:Summary}.

\section{MFCNN approach}
\label{sec:model}

So far, the matched filtering method~\cite{PhysRevD.60.022002} plays a vital role in ground-based GW detection and has achieved great success in the discoveries of many GW events.
It is based on an existing full waveform template bank which grids up parameter space sufficiently densely.
Thus, GW signals beyond the template bank cannot be easily detected.
For the first observational run of Advanced LIGO, the bank targeting compact binary systems contains $\sim$250,000 templates~\cite{roy2019effectual}.
Considering more complicated physics of GW sources, the number of templates in the bank will increase.
Although it is computationally expensive to calculate matched-filtering signal-to-noise ratios (SNRs)~\cite{helstrom2013statistical} with all templates in the bank, the process can extract features of weak signals from noisy data.
Based on this point, we construct the first layer of the MFCNN to implement matched filtering process by using a small amount $N_t$ of waveform templates as learnable weights.
Practically, the input strain data $d$ and the prepared waveform templates $h$ are first whitened as
\begin{subequations}
	\begin{align}
		\bar{d}(t) &= d(t) \, \star \, \bar{S}_n(t)\,, \\
		\bar{h}_i(t) &= h_i(t) \, \star \, \bar{S}_n(t) \,,
	\end{align}
\end{subequations}
where $\star$ denotes convolution operation and $\bar{S}_n(t) = \int^{+\infty}_{-\infty} df S_n^{-1/2}(f) e^{i2\pi ft}$ is related to the given noise power spectral density $S_n(f)$.
Then, an expected matched filter is calculated as $\bar{d}(t) \, \star \, \bar{h}_i(-t)$ with the fixed coefficient $\bar{h}_i$ ($i=1, 2, \cdots , N_t$).
Finally, analogous to the standard matched filtering approach, maximize and normalize the output of each filter as
\begin{equation}
	\bar{d}(t) \, \star \, \bar{h}_i(-t) \quad \longrightarrow \quad \max \left[ \frac{\bar{d}(t) \, \star \, \bar{h}_i(-t)}{ \sqrt{[\bar{h}_i(t) \, \star \, \bar{h}_i(-t)]|_{t=0}}} \right],
\end{equation}
which corresponds to the SNR for each template.
The output SNRs from the first matched-filtering layer capture the general features of GW signals.
The rest part of the neural network is employed as the usual convolutional neural network (CNN)~\cite{lecun1998gradient} to analyze these features.
Many studies~\cite{PhysRevLett.120.141103, PhysRevD.97.044039, George:2017pmj, PhysRevD.100.063015, PhysRevD.101.104003, krastev2020real, PhysRevD.102.063015, PhysRevD.103.024040} have indicated that CNN can capture patterns in the data, and we show that the architecture we built works as well by extracting features from the gravitational wave templates.
In the end, a Softmax function is applied to evaluate the predictions of confidence scores for searching MBHB signals, which is commonly used in the task of GW signal detection.

The MFCNN model combines the power of template matching to identify weak signals and the ability of CNN to extract features without any prior knowledge.
It does not generate triggers for a single channel, but develops instead a general convolutional neural network which can be employed afterwards to recognize coherent patterns between the matched filters.
The specific structure of the MFCNN model used in this work is similar to the Fig.1 in~\cite{PhysRevD.101.104003}.
The model constructed in~\cite{PhysRevD.101.104003} targets to search for GW signals emitted from stellar mass compact binaries and it is capable of identifying all GW events in GWTC-1~\cite{abbott2019gwtc}.
In this work, the structure of MFCNN model is adjusted to adapt to space-based GW detection.
Instead of data from LIGO interferometer channels used in~\cite{PhysRevD.101.104003}, we implement TDI observables~\cite{Tinto:2004wu} as model input channels.
Moreover, in the first layer, we set more templates and use the noise power spectral density of LISA to conduct the whitening process.
Actually, more differences between the approaches of this work and Ref.~\cite{PhysRevD.101.104003} lie in the characteristic of datasets that will be introduced in detail in Sec.~\ref{sec:Training}.

\section{Search strategy}\label{sec:strategy}

In this section, we develop our conceptual contributions to search strategy for GW signals, namely that (a) probability predictions from any models output with the Softmax function are not suited to claim statistically significant detections of gravitational waves, however, (b) they can still be useful statistics for trigger generation.
This contribution is especially important for understanding how to identify the temporal location of potential GW signals and diminish false alarms in streaming data.

Our core argument for claim (a) hinges on the fact that the negative log-likelihood cost function applied on models with the Softmax function always strongly penalizes the most active incorrect prediction, and the correct output for samples with large SNR contributes little to the learnable parameters of the model (see Sec. 6.2.2.3 in~\cite{goodfellow2016deep}).
Although certain binary classifiers can produce effective decision boundaries between signal-free and non-signal-free samples, they perform badly in ranking influential non-signal-free samples, which are not monotonically influential on SNR or other statistics~\cite{PhysRevD.101.104003}.
Consequently, the significance level obtained from model prediction with Softmax cannot be treated as valid ranking statistic for distinguishing loud triggers from fainter ones.

To substantiate (b), we highlight that by either ``repeated-modeling" or ``repeated-sampling" analysis on the short data segment, one can provide a search strategy to indicate confidence level and diminish false alarms.
For example, the ``repeated-modeling"-based multi-model stacking ensemble learning for GW detection~\cite{huerta2021accelerated,wei2021deep,PhysRevD.105.083013} can combine the output of each independent sub-model to identify GW triggers that pass a given threshold.
Intuitively, the probability of misclassification decreases with an increase in the number of sub-models.
Unlike the above ``repeated-modeling" analysis, ``repeated-sampling" method is an analysis of repeated measurements to capture the variability of a sample statistic.
Similar to how time-shifted analysis~\cite{usman2016pycbc} has been used to create a larger number of data to approximate background noise, the method, first applied in~\cite{PhysRevD.101.104003}, repeatedly analyzes varied snippets of the target data using a single model to identify the significance of candidates with frequentist interpretations.
This method takes longer data segments with significantly overlapping each neighbouring samples as input data and repeatedly estimates the background around the target short data segment.
Importantly, in detecting a signal hiding in the streaming data, a simple value comparison of continuous alerts from model prediction can help in detecting the trigger and ruling out false positive candidates.

To detect MBHB signals in streaming data, we implement the output of the MFCNN as a ranking statistic for trigger generation.
More specifically, in a real search, we adopt the ``repeated-sampling" method to search for MBHB signals in a very-long-duration dataset from the LISA Data Challenge (LDC)~\cite{ldc}.
For each input segment of a given streaming data, we choose to overlap segments by $80 \% $ in the MFCNN process.
In other words, as we continuously advance by one fifth of the length of the input, each overlapping snippet appears in five segments and is thus processed 5 times by our model.
Subsequently, we define the local-maximum trigger as 5 (or more than 5) neighbouring segments with the same values predicted by the MFCNN and surrounded by the segments with smaller values.
For record purposes, we output the centre time of each input segment and the potential coalescence time of MBHB signals.

\section{Datasets}\label{sec:Training}
\begin{figure}[htb]
    \centering
    \includegraphics[width=1\linewidth]{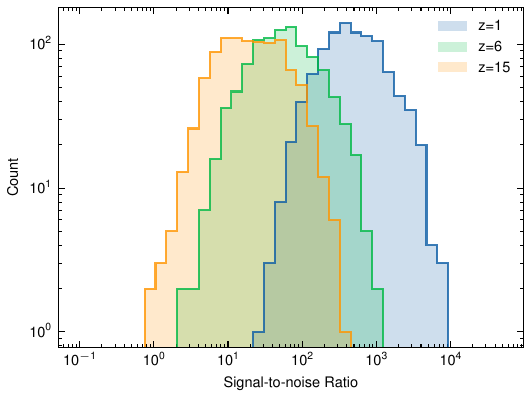}
    \caption{SNR distribution of the training and validation datasets with respect to different redshift values.}
    \label{fig:dist_SNR}
\end{figure}

We use the IMRPhenomD waveform model~\cite{PhysRevD.93.044006, PhysRevD.93.044007} to simulate 3,000 waveforms of non-precessing MBHBs at redshift $z$ between 1 and 15.
In practice, the redshift is converted to luminosity distance assuming a flat $\mathrm{\Lambda CDM}$ cosmology with $\Omega_m=0.31,\Omega_\Lambda=0.69$ and $H_0=67.74$~\cite{ade2016planck}.
More specifically, we use a logarithmic scaling to sample redshifted total mass $M_z = M(1 + z)$ in the range of $(10^{5.4}M_\odot, 10^{8}M_\odot)$ with steps of 0.013.
We also sample mass ratio $q$ in the range of $(1, 15)$ with steps of 1 ($M = m_1 + m_2$ is the total mass of the two companions).
We sample the coalescence time $3 \mathrm{d} \leq t_c \leq 365.25 \mathrm{d}$ for nearly a whole year and adopt a uniform prior over the binary's dimensionless spins $-0.9 \leq (s_{1z}, s_{2z}) \leq 0.9$ with spins aligned with the orbital angular momentum.
We assume a uniform prior over the remaining parameters (inclination angle $\iota$, ecliptic latitude $\beta$, ecliptic longitude $\lambda$, reference phase $\phi_c$ and polarization angle $\psi$), and they are sampled in the range of $\iota \in [0,\pi], \beta \in [-\pi/2,\pi/2], \lambda \in [0,2\pi], \phi_c \in [0,2\pi], \psi \in [0,\pi]$.

The Gaussian instrumental noises are generated and based on the power spectral density (PSD) stated in the LISA Science Requirement Document~\cite{lisa_sci_rs}.
The other component of the noise comes from the GW signals of Galactic binaries, although some of the signals contained in this component are detectable individually.
We simulate this kind of noise using the PSD estimated from LDC noiseless data~\cite{ldc} which contains simulated waveforms from 30 million Galactic binaries.
The noise generated in this way is both Gaussian and stationary.
Though, in fact, the Galactic signals should be treated as non-Gaussian and non-stationary noises modulated by the motion of the detector~\cite{Adams:2010vc, cornish2021low}.

For LISA, the TDI technique is applied to suppress the laser frequency noise~\cite{PhysRevD.65.102002, Tinto:2002de, PhysRevD.66.122002, Tinto:2003uk, PhysRevD.68.061303, Tinto:2003vj, Cornish:2003tz}.
We choose the uncorrelated TDI observables A and E~\cite{PhysRevD.66.122002} as two channels of the data passed through our neural network.
Using the LDC's code provided in challenge-1 (LDC-1)~\cite{ldcmanual001}, we generate data using TDI-1.0~\cite{Vallisneri:2004bn}.

As a tradition in training deep learning models, our data is divided into training and validation sets.
The validation data set contains the same data distribution as the training data set and allows for an unbiased evaluation of a model fit on the training data set while tuning the model's hyperparameters.
To develop a classification model, we split the training datasets into two categories, one containing MBHB signals and additive Gaussian noise, and the other containing Gaussian noise alone.
In the training phase, we set the input size of the MFCNN as $16,384$ and the sampling rate as $1/15$ Hz.
For the templates utilized in the first layer of the MFCNN, we sample the logarithm of $M_z$ uniformly between $(10^{5.4}M_\odot, 10^{8}M_\odot)$ for 50 equal-mass MBHB sources at z=3 for convenience.
Unlimited amounts of training and validation dataset that combines noise and modeled waveform are generated in our experiments to overcome overfitting.
The data is augmented by randomly shifting the signals within the input segment of 50-90\%.
Fig.~\ref{fig:dist_SNR} shows the distribution of SNR for the training and validation datasets with respect to different redshift $z$ values.
The training process takes 10 hours on a NVIDIA GeForce RTX 2080 GPU with 8GB of memory.
Conventionally, we output the optimal model which works quite well in searching for MBHB signals with IMRPhenomD waveforms.

\section{Results}\label{sec:result}
\subsection{Generalization test}
\begin{figure}[htb]
    \centering
    \includegraphics[width=1\linewidth]{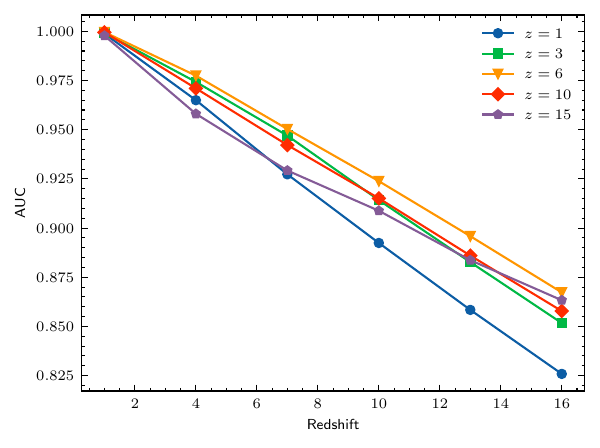}
    \caption{AUC of the MFCNN models as a function of redshifts. The testing datasets are generated by IMRPhenomD model. The colored lines denote the five MFCNN models trained on datasets with $z = 1, 3, 6, 10, 15$, respectively.}
    \label{fig:model_auc_z}
\end{figure}

In this section, we employ the receiver operating characteristics (ROC) analysis~\cite{egan1975signal, fawcett2006introduction} to visualize the performance of the MFCNN model on the testing dataset.
The ROC curve is a graphical plot that depicts the changes in true positive rate (TPR) and false positive rate (FPR)~\cite{fawcett2006introduction} as the discrimination threshold is varied.
To compare classifier performance, the area under the ROC curve (AUC)~\cite{bradley1997use, hanley1982meaning, fawcett2006introduction} has long been used to quantify the ROC performance as a single scalar value varying between 0 and 1.
The AUC is a threshold-free metric capable of measuring the overall performance of binary classifiers.
The closer the AUC to 1, the better the classifier.


We test the MFCNN model on datasets which assume independently identically distributed about the training set.
Each testing dataset with various redshift $z$ consists of 3,000 positive samples (with MBHB signals) and 3,000 negative samples (without MBHB signals).
We generate waveforms of MBHB signals using the IMRPhenomD model, using the same priors as those used in training phase.
To ensure the training samples will not appear in test dataset, We shift the grid of $M_z$ and $q$ from the values in training data.
Specifically, we use a logarithmic scaling to sample redshifted total mass $M_z = M(1 + z)$ in the range of $(10^{5.4 + 0.0065}M_\odot, 10^{8 + 0.0065}M_\odot)$ with steps of 0.013 and sample mass ratio $q$ in the range of $(1 + 0.5, 15 + 0.5)$ with steps of 1.
The priors of the other parameters are the same in the training phase.

\begin{figure}[htb]
\centering
\includegraphics[width=1\linewidth]{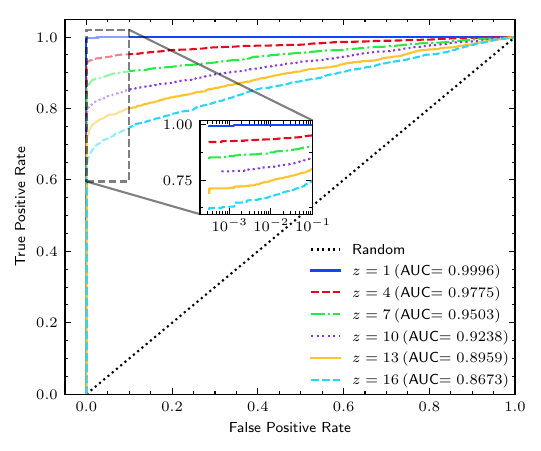}
\caption{ROC analysis for the MFCNN model trained with $z=6$. The redshifts of the testing datasets are selected to be $1, 4, 7, 10, 13, 16$. The black dotted line denotes the line of random classifier. The grey dashed box indicates the region of the inset.}
\label{fig:model_z6}
\end{figure}

We show the performances on testing datasets with various redshifts in Fig.~\ref{fig:model_auc_z}.
It is consistent that, as the redshift of test dataset increases (SNR decrease as seen in Fig.~\ref{fig:dist_SNR}), the classification ability of MFCNN models declines.
We found that our MFCNN model attains optimal performance in AUC as we shift the redshift between $z=1$ and $z=16$ for testing while the model trained on $z=6$ achieved best performance among the others.
Here, we will use this model for the following test.
In Fig.~\ref{fig:model_z6}, we also vary the threshold of prediction from 0 to 1 to show the ROC curves for each test dataset.
As we vary the observing limit from $z=1$ to $z=16$ for $\text{FPR}=10^{-3}$, the $\text{TPR}$ decrease from 0.997 to 0.627. 
This shows our model achieves a high level of sensitivity and can capture the distinctive waveform features of MBHBs in highly noisy environments.

In reality, MBHB signals have richer features than the approximate waveform template.
The IMRPhenomD waveform only considers aligned-spin black hole binaries with circular orbits.
However, more complicated evolution of MBHBs in LISA band need to be considered, such as residual eccentricity~\cite{Sesana:2010qb} and precession of binary's orbital plane~\cite{PhysRevLett.70.2984, Apostolatos:1994mx, Vecchio:2003tn}, as MBHB waveforms will be modulated due to these facts.
This is necessary for the neural network to work well on MBHB waveforms beyond the training dataset.

Therefore, we generate 3 additional test datasets based on SEOBNRv4~\cite{PhysRevD.95.044028}, SEOBNRE~\cite{Cao:2017ndf, Liu:2019jpg, liu2021highermultipole} and SEOBNRv4P~\cite{Pan:2013rra, Babak:2016tgq, PhysRevD.102.044055} waveform family.
The SEOBNRv4 dataset describes the same binary system as IMRPhenomD but differs in modelling implementation.
The SEOBNRE dataset contains the GW waveforms of eccentric MBHBs, and we sample parameters of eccentricity uniformly in the range of $(0, 0.3)$.
The SEOBNRv4P dataset captures the precession-modulated waveforms, in which the components' dimensionless spins $(s_{1x}, s_{1y}, s_{1z})$ and $(s_{2x}, s_{2y}, s_{2z})$ of binary system are uniformly sampled in the range of $(-0.9, 0.9)$.

\begin{figure}[!htbp]
\centering
\includegraphics[width=1\linewidth]{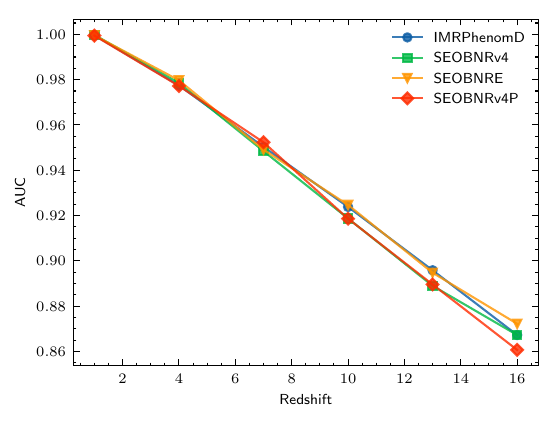}
\caption{Performance of the MFCNN model on different waveform families. The testing datasets with various redshifts are generated by IMRPhenomD, SEOBNRv4, SEOBNRE and SEOBNRv4P. The AUC~\cite{bradley1997use, hanley1982meaning, fawcett2006introduction} is a commonly used metric to compare different classification algorithms.}
\label{fig:auc_acc}
\end{figure}

In Fig.~\ref{fig:auc_acc}, we test the MFCNN model on the various datasets generated by the four waveform families.
Compared to the IMRPhenomD waveform used for training, the model's performance on modulated waveform families is fairly similar throughout a wide range of redshift evaluated here.
Even if the signal-to-noise ratio of a distant waveform event is very low, such as $z=16$, the model can generalize to various waveform variations.
Moreover, it turns out that our model has the nice ability to robustly extrapolate beyond the representations of our training region.
This implies that our model may have the power to search real LISA data in the future for MBHB signals beyond the theoretical templates.

\subsection{Sangria dataset} \label{subsec:Sangria}

To assess how it handles more realistic data, we use the Sangria dataset from LDC-2~\cite{ldc} to evaluate the trigger generation performance of our MFCNN model.
The dataset covers approximately a year of simulated LISA data and contains simulated waveforms of 30 million Galactic binaries, 17 verification Galactic binaries and 15 coalescing MBHBs.
We down-sample the dataset to $1 / 15$ Hz for consistency.
We then divide the dataset into overlapping segments corresponding to the input size of our model.
We choose to overlap the segments by $80 \% $ as discussed in Sec.~\ref{sec:strategy} and then pass the data through the MFCNN model.
Processing all the segments with our MFCNN model produces a sequence of predictions which we use in further analysis.
Specifically, the find\_peaks algorithm provided by SciPy~\cite{2020SciPy-NMeth} is used to detect a peak plateau on the probability of the positive class predicted by our model.
Based on our search strategy in Sec.~\ref{sec:strategy}, we use the previously described local-maximum triggers to identify MBHB coalescences, and we output the centre time of the segments that are potential merger locations.
Notably, it takes only several seconds for the model to analyse a year long dataset on our device.

\begin{figure*}[!htbp]
\centering
\includegraphics[width=\textwidth]{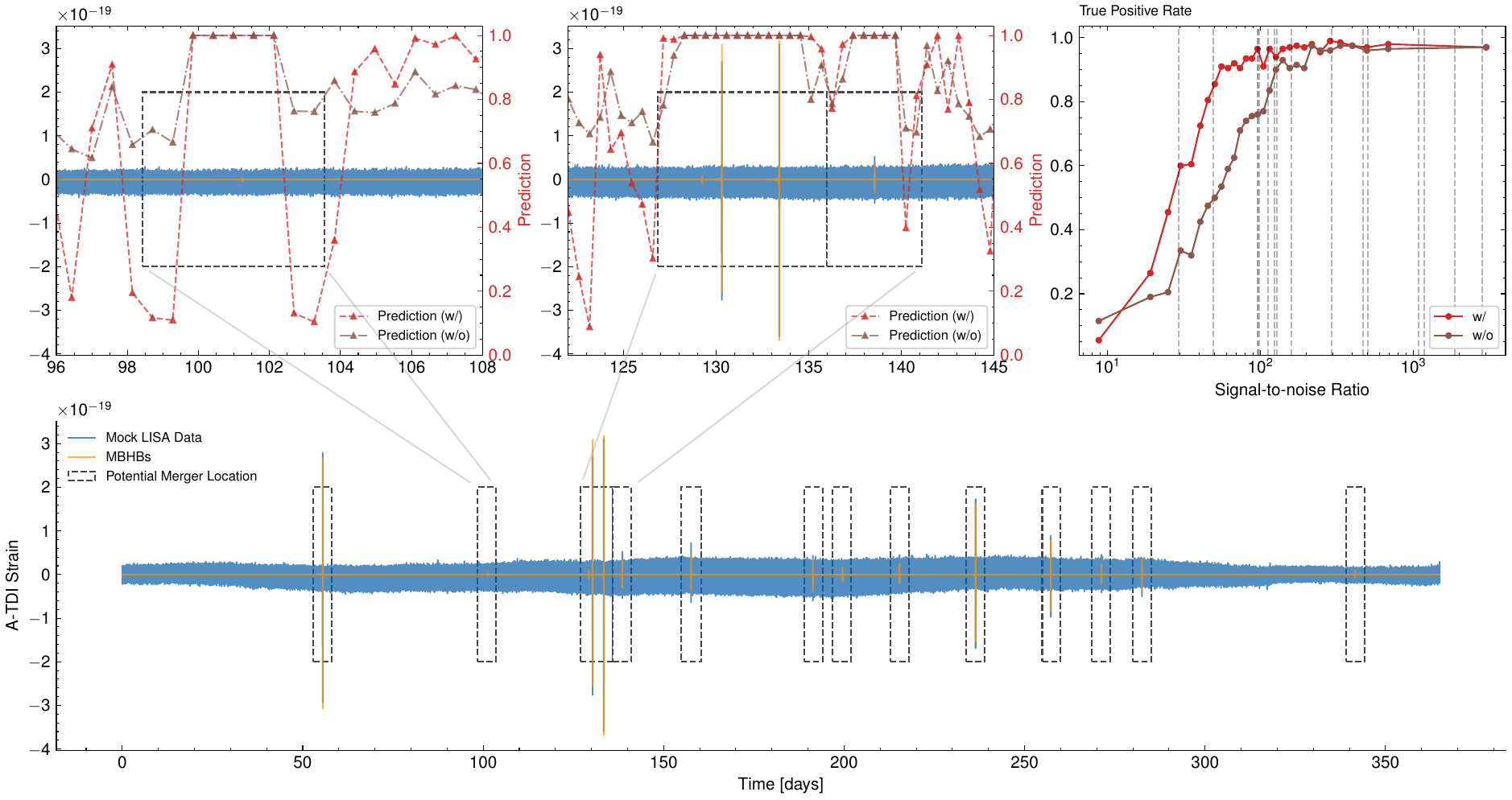}
\caption{Prediction of the MFCNN model on 1-year LISA data simulated by the LDC group. In the bottom panel, the blue line is the A-TDI data, the orange line denotes the 15 MBHB signals which are clearly visible and the dashed boxes denote the range of coalescence time predicted by our model. Two zoomed regions are shown in the first and second upper panel. The red/brown triangle indicates the prediction from the MFCNN model. The notion of ``w/" and ``w/o" means the model trained with and without mixed confusion noise. The upper right panel shows true positive rate (the fraction of 6,000 injected MBHB signals correctly identified) versus the SNR of the signals. The dashed line corresponds to SNR of the 15 MBHB signals respectively.
}
\label{fig:LDC_test}
\end{figure*}

As the main result of this work, we demonstrate the predictive ability of the MFCNN model on the Sangria dataset in Fig.~\ref{fig:LDC_test}.
For this analysis, we used two types of MFCNN for MBHB classification: models trained with and without mixed confusion noise.
Both models, it turns out, locate all 15 MBHB mergers to within a short segment.
Particularly, on the upper left of Fig.~\ref{fig:LDC_test}, even the MBHB signal with the smallest amplitude can be clearly recognized by our MFCNN models. 
Each trigger implies that the range of coalescence time is within 5.12 days.
The target MBHB mergers are found right in the middle of this range.
We find that this is always the case.
Notably, we can also achieve the desired output even when multiple MBHBs merger are close together.
Examples of this are shown in the upper middle panel of Fig.~\ref{fig:LDC_test}.

Although GW signals from Galactic binaries in LDC's 1-year data are generally considered non-Gaussian and non-stationary noise~\cite{Adams:2010vc, cornish2021low}, the MFCNN models are simply trained on datasets with Gaussian and stationary noise regardless of whether it is mixed with or without the estimated confusion noise.
In addition, we setup, for simplicity, a static configuration with a TDI-1.0 response~\cite{ldcmanual001} to generate our training data and measure the sensitivity (see caption of Fig.~\ref{fig:LDC_test} for further explanations) of the MFCNN models as shown in upper right panel of Fig.~\ref{fig:LDC_test}.
However, considering the complexity of the spacecraft motion, we perform a thorough analysis on the Sangria dataset coded by LISANode~\cite{Bayle:2018hnm} with a TDI-1.5 response~\cite{ldcmanual002}, which applies to a rigid but rotating configuration.
Although the code and TDI version are inconsistent with our training data, we find that our model still can recognize all MBHB signals and reports no false alarms.
It implies that the MFCNN model shows a robust resistance to the dynamic modulation of space-based GW detectors and has the potential to capture the general features of the waveform response.

\section{Summary and discussions}
\label{sec:Summary}
In this study, we demonstrate that the deep learning method, when applied to LISA data, is capable of searching for MBHB coalescences.
We further employ the MFCNN model with a small number of templates to analyse and output predictions on a year long Sangria dataset within ten seconds.
Our model can identify all 15 MBHB mergers with no false alarms and locate each merger to data segments as short as 5.12 days long.
These results lay the foundation for accelerating the detection and forecasting the mergers of MBHBs to enable the observation of EM mission emerging after the MBHB coalescence.
By building a neural network capable of rapidly searching for and counting MBHBs, we answer a fundamental question regarding the applicability of neural networks to LISA data analysis.

In practice, reliable searches for MBHB signals in streaming data are strongly affected by non-Gaussian and non-stationary noise, such as from unresolved Galactic binaries.
To account for the overlap between signals, global-fit approaches~\cite{PhysRevD.72.043005, Littenberg:2020bxy,littenberg2023prototype} are adapted for space-based GW detectors to model all resolvable signals and instrument noise.
The MFCNN-based analysis provides the number of sources and the time of coalescences,
which are useful for the subsequent global-fit analysis.
According to the number of sources we can fix the dimension of the parameter space in the subsequent global fit.
If the number of sources is unknown, model selection has to be employed to determine the number of sources.
In the conventional approach to model selection we need to calculate the Bayes factor between competing models of different dimension.
Compared to the fixed-dimension analysis with the help of MFCNN,
the global fit including model selection is several times computationally costly.
As mentioned in Sec.V our model takes several seconds to analysis a year of data.
However, the global fit for a MBHB with a year of data usually takes several hours on a multi-core processor.
Actually the computational cost of parameter estimation for MBHB signals is affected by the SNR and the dimension of the parameter space.
Of cource, for different global-fit algorithms, the MFCNN model may bring different improvements.
Moreover, the time of coalescences provided by MFCNN can be used to determine or optimize the observation time needed to achieve a desired result.
Our analysis represents a starting point for applying a neural network trained on Gaussian and stationary background noise to realistic non-Gaussian and non-stationary data.
Due to the large dimensionality of the data characteristics needing modelling, there exists the potential for neural networks to exceed the sensitivity of existing Bayesian analysis in regard to real data.


In real-world tasks, the actual noise in recordings will need to be estimated simultaneously along with the various potential signals.
However, in general, one can obtain information of the noise from a specific TDI observable in which the signals are greatly depressed~\cite{PhysRevD.66.122002}.
Nevertheless, as can be seen in Fig.~\ref{fig:LDC_test}, it makes little difference on the Sangria dataset whether or not the MFCNN model is trained using confusion noise.
This approach is essential to ensuring that neural networks properly characterize the non-Gaussian and non-stationary nature of realistic detector noise experienced by observatories.
We further test the two models with additional injected signals and found that there is a slight difference in performance for signals with SNR $<100$.
As a result, once fully trained, the model can be used for real-time MBHB searches or even to achieve early warning in practice~\cite{PhysRevD.103.063034,Wei_2021,PhysRevD.104.062004}.

In this work, we present a model with robust sensitivity to numerous GW sources and modulation of MBHB waveform family.
As we mentioned in the case of the Sangria dataset, the generalization ability of the supervised learning approach can be extended to various TDI configurations and can also be greatly useful for future space-based GW detectors.
There will definitely be differences between the real data and the simulated data, such as the existence of glitches, non-stationary instrumental noise and data gaps (see discussion in~\cite{cornish2021low}), so the performance and robustness of our method on realistic SNRs needs to be assessed and understood.
However, more detailed investigations, such as the cosmic population of MBHBs, are beyond the scope of the present work.
The analysis described here is an initial application of the deep learning approach to searching for MBHBs via space-based detection.
Note that, the MFCNN model is trained with GW waveforms containing the merger phase.
In the future, it is more inclined to train a deep learning model which is capable of identifying MBHB signals at the early inspiral stage for multi-messenger astronomy.
The approach developed in this work can provide a potential application for LISA-like observatories to identify and localize MBHBs during the inspiral phase allowing time for detailed planning and coordination of joint multi-messenger observations.






\begin{acknowledgments}
We thank Zhoujian Cao for his helpful comments and discussions.
CL would like to thank Stanislav Babak for help in the use of the LDC code and datasets.
WHR would like to thank Pengfei Zhou for discussion on CNN and Xiaolin Liu for discussion on SEOBNRE waveform model.
We thank the Peng Cheng Laboratory (PCL) Cloud Brain for computation support.
This work is supported in part by the National Key Research and Development Program of China Grant No. 2020YFC2201501, in part by the National Natural Science Foundation of China under Grant No. 12075297 and No. 12235019.
The authors would like to acknowledge the work of the LDC group.
For this study, both the LDC software and datasets were used~\cite{ldc}.
PyCBC~\cite{alex_nitz_2021_5347736} and LALSuite~\cite{lalsuite} are also used to generate gravitational strains of coalescing MBHBs.
Plots are generated by Matplotlib~\cite{Hunter:2007,thomas_a_caswell_2021_5194481}.
The implementation of the MFCNN model is coded based on PyTorch~\cite{NEURIPS2019_9015}.

\end{acknowledgments}

\bibliography{main}

\end{document}